\documentclass[prx,twocolumn,10pt]{revtex4-1}

\usepackage{amsmath,amssymb,amsfonts}
\usepackage{ulem}

\usepackage{graphicx}
\usepackage{epstopdf}
\usepackage{xcolor}
\usepackage[utf8]{inputenc}
\usepackage[T1]{fontenc}
\DeclareGraphicsExtensions{.eps}

\definecolor{darkblue}{rgb}{0, 0, 0.8}
\usepackage[colorlinks=true, breaklinks=true, linkcolor=darkblue, citecolor=darkblue, urlcolor=darkblue]{hyperref}
\newcommand{\doilink}[2]{\href{http://dx.doi.org/#1}{#2}}

\begin{document}
\title{Fabrication and characterization of super-polished wedged borosilicate nano-cells.}

\author{T.Peyrot}
\author{Ch. Beurthe}
\author{S. Coumar}
\author{M. Roulliay}
\author{K. Perronet}
\author{P. Bonnay}
\author{C.S. Adams}
\author{A. Browaeys}
\author{Y.R.P. Sortais}




\begin{abstract}
We report on the fabrication of an all-glass vapor cell with a thickness varying linearly between (exactly) 0 and $\sim 1$ $\mu$m. The cell is made in Borofloat glass that allows super polish roughness, a full optical bonding assembling and easy filling with alkali vapors. We detail the challenging manufacture steps and present experimental spectra resulting from off-axis fluorescence and transmission spectroscopy of the Cesium D1 line. The very small surface roughness of $1${\AA} rms is promising to investigate the atom-surface interaction or to minimize parasite stray light.
\end{abstract}

\maketitle
\section{Introduction}
Miniaturization of devices based on atomic vapor has set the stage for the development of various compact light-matter interfaces\,\cite{Wasilewski2010,Petelski2003}. However, to reach high precision performances required to be used as atomic sensors\,\cite{Knape2005,Schaeffer2015}, integrated optical chips immersed in alkali-vapors demand meticulous characterization. In this perspective, vapor nano-systems\,\cite{Sarkisyan2004,Sarkisyan2001,Low2018} have recently been used to clarify fundamental problems in optics such as the role of dipole interactions in resonant light scattering\,\cite{Peyrot2018,Low2018}, control the coherent excitation of Rydberg atoms\,\cite{Pfau2010}, or evidence the non-local response of an atomic gas\,\cite{PeyrotArxiv}. Another issue inherent to the system size reduction, is the growing influence of the long-range atom-surface interaction, that triggered the development of new glass nano-cells\,\cite{Baluktsian2010,Whittaker2015,Whittaker2017}. Also, by combining short repulsive and long attractive potential, the possibility to trap atoms in bound states has theoretically been predicted\,\cite{Lima2000} and this ability could  make way to hybrid nanoscale atom-surface meta-materials\,\cite{Whittaker2014}.

\begin{figure}[b!]
\includegraphics[width=\linewidth]{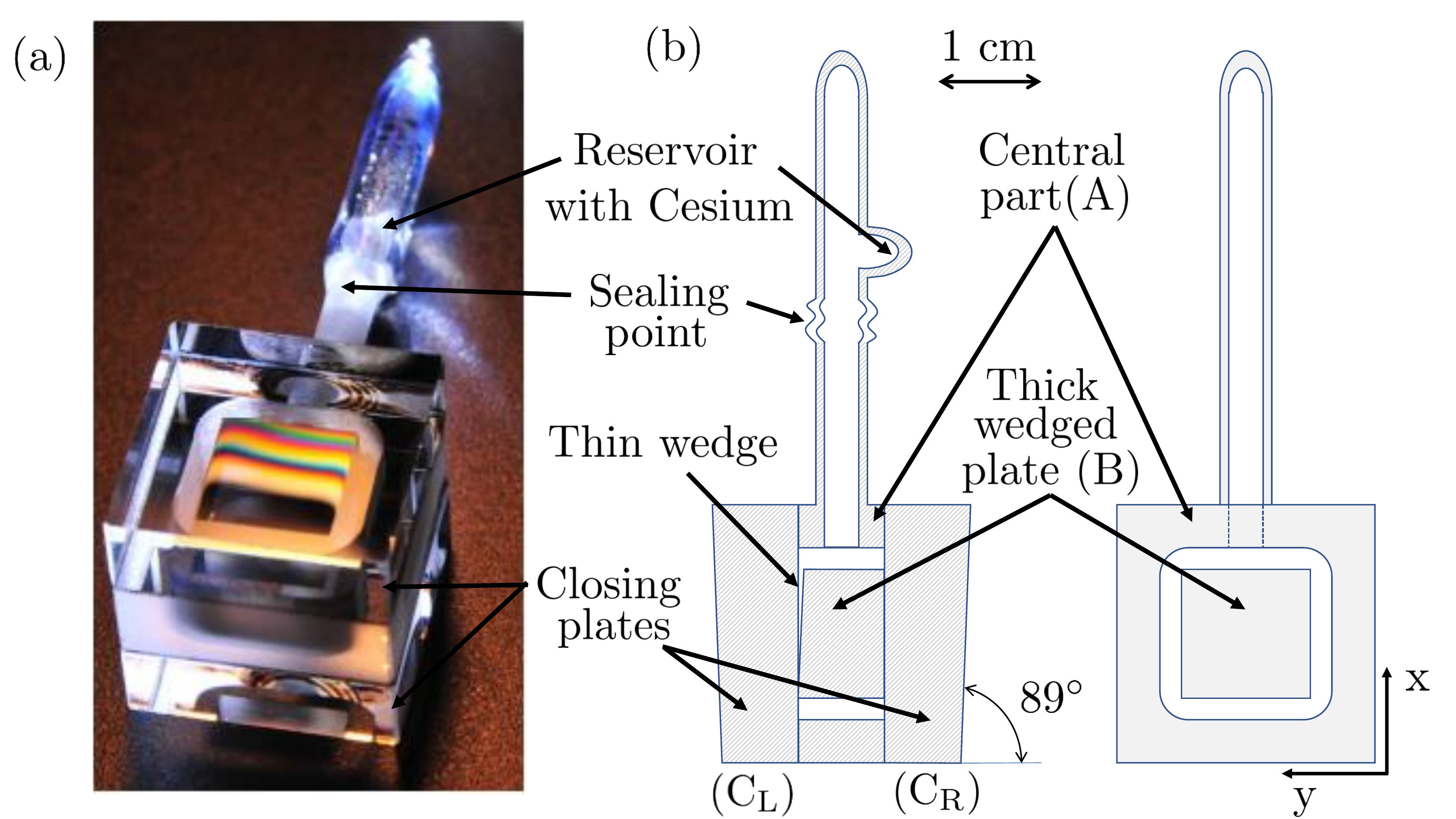}
\caption{(a)Photograph of the cell. The white light color fringes reflect the thickness variations in the wedge of the nano-cell. (b) Schematic of the cell.}
\label{fig:Fig1}
\end{figure}
In this article, we report on the fabrication of a glue-free all-glass nano-cell with a thickness varying linearly between (exactly) $0$ and $\sim1\,\mu$m, and with a surface roughness of $1${\AA} rms. Filled with Cesium (Cs), the cell was initially built for the interrogation of small atomic vapor layers of alkali and investigate the hypothetical existence of an atom-surface bound state that would depend on the surface corrugation\,\cite{Vargas1998}. Developing surfaces with ultra-low surface roughness is also an asset to reduce parasite scattering limiting signal-to-noise ratio of light matter interaction with nanometric atomic ensembles. Besides, the super-polish surfaces presented here can be ideal candidates for the fabrication of glass micro-cavities for QED experiments \cite{Barrett2011}. Therefore, the techniques that we present may have a large panel of applications and find audience in diverse domains such as surface science optics and atomic physics. We will first describe the cell fabrication process and how we filled it with Cs. We subsequently present the methods to check the requirements in terms of surface roughness and cell thickness. Finally we bring forward promising examples of spectra in transmission and off-axis fluorescence of the Cs D1 line.

\section{Fabrication and characterization}
\label{sec:Fabrication and characterization}

The cell was fabricated at Laboratoire Charles Fabry of Institut d'Optique (Palaiseau, France) and filled with a cesium vapor at the Laboratoire GEPI of Observatoire de Paris (Paris, France). It is made of four parts that are assembled by optical contact bonding  leading to a monolithic ensemble (see Fig.~\ref{fig:Fig1}a). Using a unique material for all parts avoids differential thermal expansion that could damage the optical bonding. Borofloat glass has been chosen for its good optical properties in the visible and near-infrared spectrum, low cost and  facility to be super-polished in comparison to other materials used in previous cells (sapphire for instance). This glass was chosen also to facilitate the sealing to the Pyrex side arm that will  contain the Cs reservoir, as the thermal expansion coefficients and the softening temperatures of both glasses are similar. However, unlike sapphire, and similar to fused silica, it reacts with alkali at temperature exceeding $\sim 200^\circ$C. The fabricated cell is therefore a good candidate for operation at moderate temperatures. The central part (A) (see Fig.~\ref{fig:Fig1}b) is machined using boring-bits so as to let a $6$\,mm external diameter, $4$\,mm internal diameter and $25$\,mm long tube protrude from the front face to allow for connection to a Pyrex loading manifold containing the Cs reservoir. Glass is then removed from the inside of part (A) using milling-bits. In this hollow piece, a thick plate (B), carefully angled on one side is introduced. The cell is closed by two thick plates (C$_L$) and (C$_R$) wedged by $1^{\circ}$ to dismiss unwanted reflections. Plate (B) is optically contacted on one side on plate (C$_R$). On the other side, the gap formed by the angle between plate (B) and plate (C$_L$) forms the nano-enclosure and its realization is therefore crucial. The two main manufacturing difficulties that we will detail are: (i) the control of the nano-gap thickness, (ii) the polishing of the different surfaces to reach excellent surface roughness.

\begin{figure}
\includegraphics[width=\columnwidth]{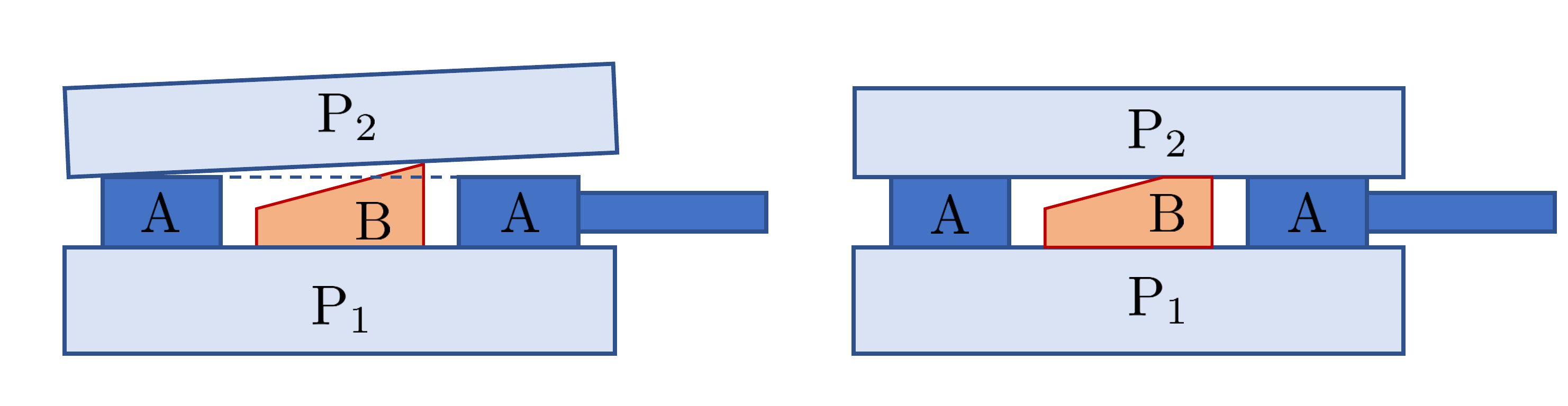}
\caption{Procedure to realize a thin wedge with thickness varying between (exactly) $0$ and $900$\,nm. Parts (A) and (B) are optically contacted on a parallel plate (P$_1$). Left : the thickness of part (B) first exceeds that of part (A) by $300$\,nm. The flat etalon P$_2$ therefore sits on the ridge of part (B), leading to interference fringes (between part (A) and the flat etalon) with different white light colours on each side of the ridge. Right : after flattening the ridge and equalizing the heights of part (A) and (B), the flat etalon sits equally on parts (A) and (B), leading to equal colour fringes.}
\label{fig:Fig2}
\end{figure}

The thin wedge of plate (B) is realized by polishing iterations and controls of the wedge thickness and flatness between two polishing steps. This is done interferometrically, using a He-Ne laser and a Fizeau interferometer. The wedge is realized such that $4$ fringes appear in the interferogram, parallel to the $y$ axis (see Fig.~\ref{fig:Fig1}) and with equal interfringe spacing. This corresponds to a thickness variation of $1.2\,\mu$m. Also, the absolute thickness is controlled such that, along its thickest edge, the plate thickness exceeds the thickness of plate (A) by $\sim 300$\,nm (see Fig.~\ref{fig:Fig2}a), forming a ridge that will eventually be removed. This is controlled mechanically, using an electronic depth gauge with a resolution of $100$\,nm. 
Parts A and B are then optically contacted on a parallel plate (P$_1$) and polished simultaneously so as to remove the $300$\,nm thick ridge and bring parts (A) and (B) to equal height for closing purpose. This height equalization is controlled interferometrically, using a flat etalon (P$_2$) as shown in Fig.~\ref{fig:Fig2}. This procedure ensures that final assembly brings the closing plate (C$_L$) in optical contact with both parts (A) and (B). The wedge thickness therefore varies between (exactly) $0$ and about $900$\,nm.

\begin{figure}
\includegraphics[width=\columnwidth]{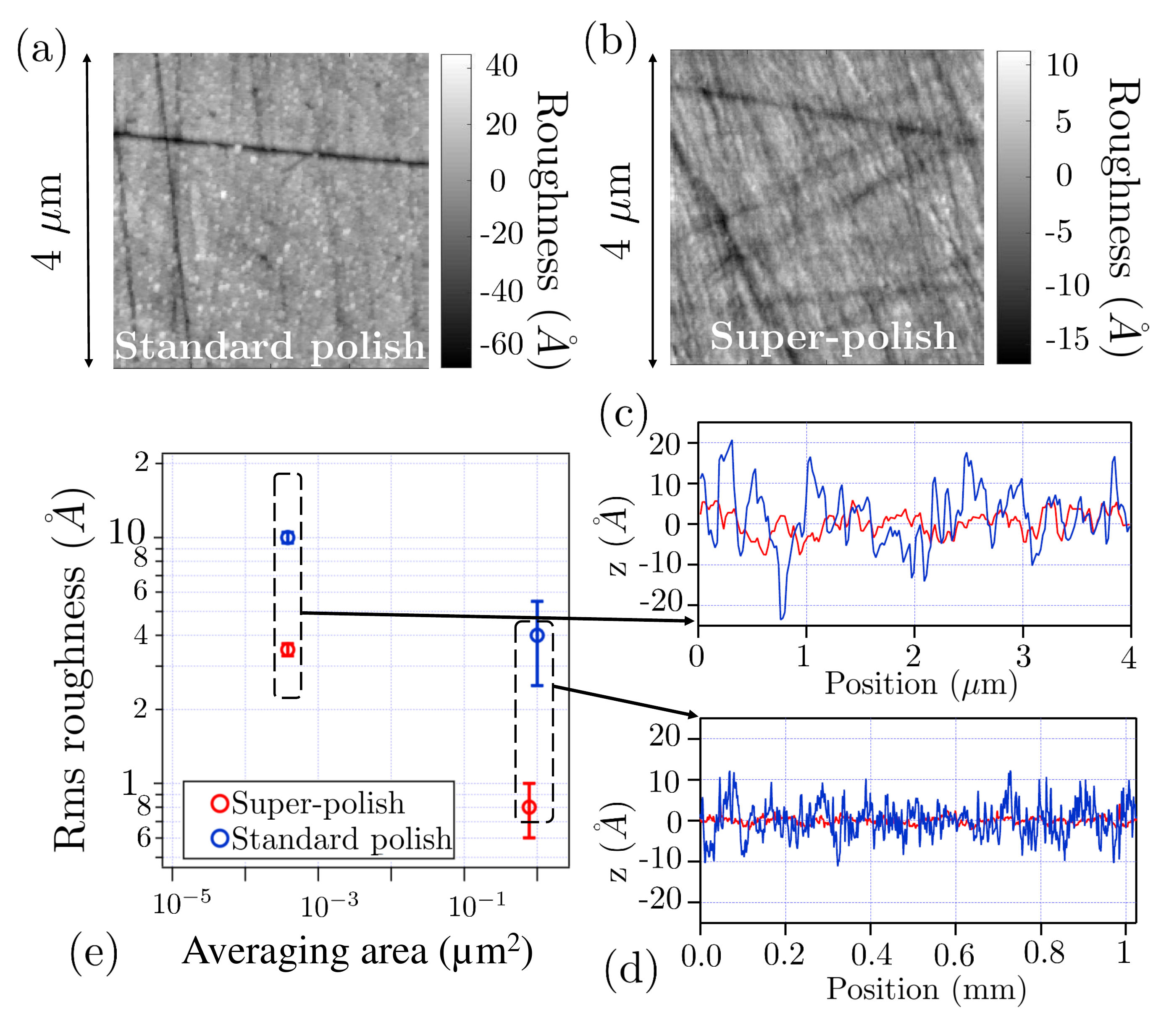}
\caption{(Top) AFM imaging of (a) a standard polish, and (b) a super-polished surface of our cell. (Bottom) Roughness profiles acquired with (c) the AFM and a spatial resolution of $20$\,nm and (d) the optical heterodyne profiler (spatial resolution : $1\,\mu$m). Red (blue) traces correspond respectively to the super-polish (standard polish) surfaces. (e) Rms roughness versus averaging area.}
\label{fig:Fig3}
\end{figure}

To realize surfaces with very low roughness such as those inside the thin wedge, we first grind them finely using alumina abrasives and a brass grinding wheel. The surfaces are then manually polished on a pitched wheel using an aqueous solution of rare earth oxide abrasives with fine particle size ($<1\,\mu$m). The final polish is performed with an increasingly diluted solution, leading to a super-polish with surface roughness of $1${\AA} rms or less. Figure~\ref{fig:Fig3} shows a typical roughness profile, measured using an optical heterodyne profiler (ZYGO 5500) with a sensitivity of $0.2${\AA} rms~\cite{Sommargren1981}. The spatial resolution of the profiler is $1\,\mu$m, on the order of the distance travelled by the atomic dipoles of the vapor before they reach steady-state (due to collisions inside the gas or radiative decay). To investigate atom-surface interactions over shorter travelling distances and account for the transient response of the vapor to finer details of the surface~\cite{PeyrotArxiv}, we characterize the surface on a smaller scale. To do so, we also acquired images of the super-polished surfaces using an atomic force microscope (AFM) in tapping mode, equipped with a Silicium tip (n-type)~\cite{AFM_model}. The spatial resolution of the AFM imaging is given by the radius of the tip, typically $10$\,nm or less. Figure~\ref{fig:Fig3} compares typical results obtained for surfaces with a super-polish and a standard polish, for which abrasives are rougher and we stopped the polishing procedure before the final dilution step. The roughness is lower for a super-polish than a standard polish, as expected, and it decreases for lesser spatial resolutions as surface height fluctuations are better averaged~\cite{Polishing}.

All parts of the cell are cleaned with alcohol prior to final assembly by optical contact bonding (no glue is used). This is done at room temperature by pressing lightly onto the parts to be bonded until white light color fringes disappear and are replaced by a uniform dark fringe where the two parts are in contact. The only requirements are the absence of dust and the complementarity of the two surfaces to be bonded : in our case, we found that a flatness difference of $30$\,nm rms or less was sufficient to achieve a stress-resistant optical bonding.

Once the cell is assembled we connected it to a turbo-molecular pumping stage via a Pyrex loading manifold, and baked the cell-manifold assembly at $350-400^{\circ}$C for 4 days whilst pumping. Cesium was then transferred, using a low temperature flame, from a cesium ampoule with breakable seal to the cell reservoir. 


\begin{figure}
\includegraphics[width=\columnwidth]{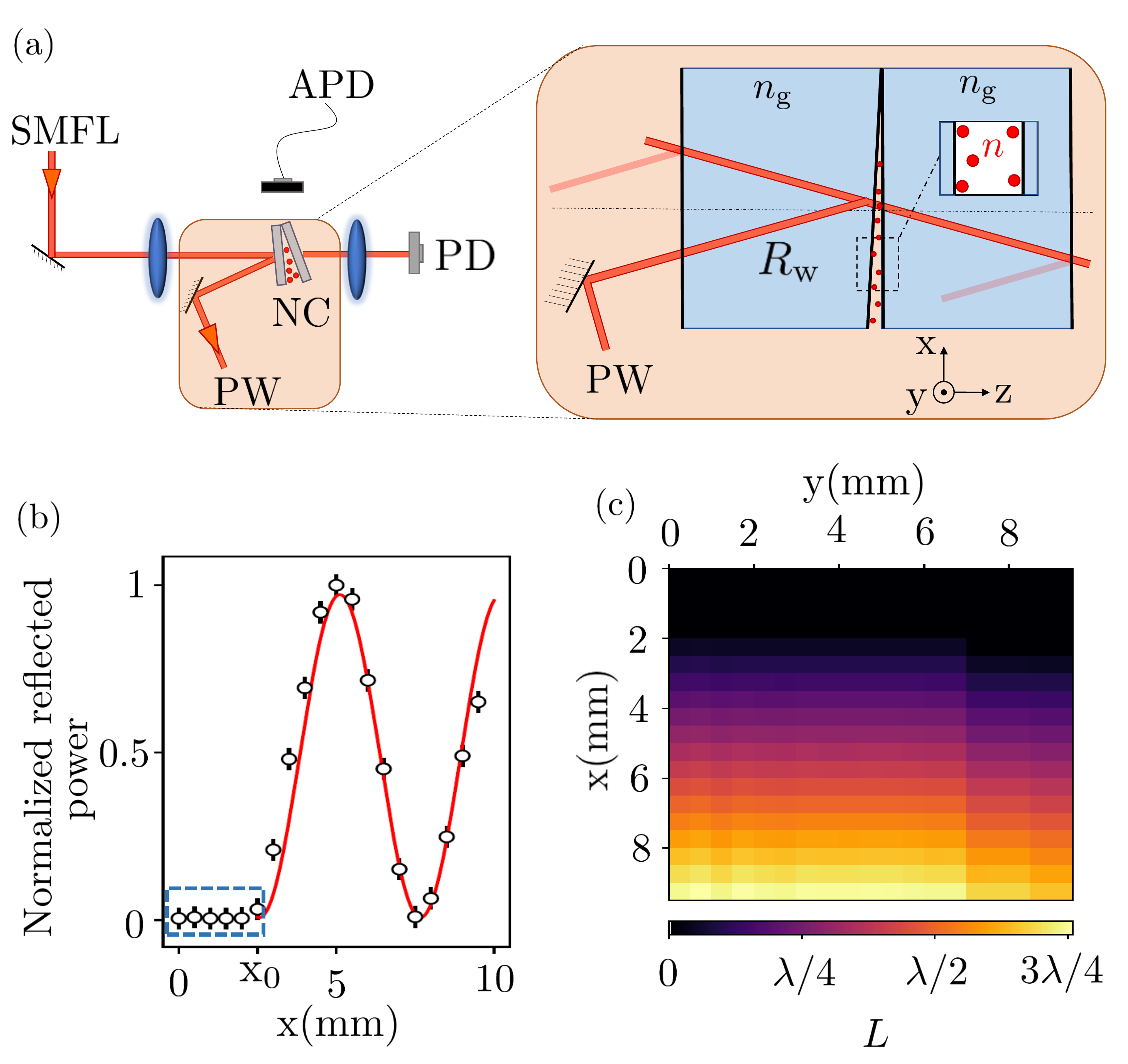}
\caption{(a) Schematic of the experiment. SMFL : single-mode fiber laser; PD : photodiode; APD : avalanche photodiode; NC : nano-cell, PW : power-meter. (b) Reflected interference signal $R_{\rm w}(x)$ (normalized to the signal obtained at a bright fringe). Black circles :  data acquired for $y=5$\,mm. Solid red line : fit by the Fabry-Perot model. For $x<x_0$ (dashed blue rectangle) the wedge thickness is exactly zero since parts (B) and (C$_{\rm L}$) are optically bonded (see Fig.~\ref{fig:Fig1}b). (c) 2D-mapping of the wedge thickness (resolution along $x$ and $y$ : $0.5$\,mm).}
\label{fig:Fig4}
\end{figure}

We now provide more information about our \textit{in situ} measurement of the wedge thickness. We access the local absolute thickness $L(x,y)$ using the interferometric technique described in Ref.~\cite{Jahier2000}. The reflected intensity $R_{\rm w}$ resulting from the interference on the thin wedge (see Fig.~\ref{fig:Fig4}a) tells about the local thickness: by scanning the laser position along the $x$ axis of the cell, from the optical contact to the thickest part of the wedge, we record on a powermeter the interferometric profile shown in Fig.~\ref{fig:Fig4}b. The optical contact is reached for $x\leq x_0$. For $x>x_0$ and a given value of $y$, the data are very well fitted by the Fabry-Perot model :

\begin{equation}\label{eq:Fabry_Perot}
R_{\rm w}(x)\propto \frac{F\sin^2(\phi(x)/2)}{1+F\sin^2(\phi(x)/2)},
\end{equation}
with $F=4r^2/(1-r^2)^2=(2\mathcal{F}/\pi)^2$ ($\mathcal{F}$ is the finesse of the wedge etalon), $r=(n_{\rm g}-n)/(n_{\rm g}+n)$, the reflectivity (in field amplitude) of the glass-to-vapor interface, $n_{\rm g}$ the glass refractive index, $n$, the refractive index of the vapor, and $\phi(x)=4\pi L(x)/\lambda$, the cumulative phase over one round-trip of the laser inside the wedge. Here, we assume that the wedge thickness varies linearly with $x$ for $x>x_0$ and the only free parameter of the fit is therefore the wedge angle, $\alpha(y)=L(x,y)/(x-x_0)$. Also, we perform this interferometric mapping of the thickness at very low vapor pressure and with laser light far detuned from the atomic resonances. The vapor index $n$ can therefore be assimilated to unity. Figure~\ref{fig:Fig4}c shows a 2D map of $L(x,y)$. The uncertainty on each point is $\pm5$\,nm, limited by the finite size of the laser beam (waist radius $w_{0}\simeq 40\,\mu$m).

%
%


\begin{figure}
\centering
\includegraphics[width=\columnwidth]{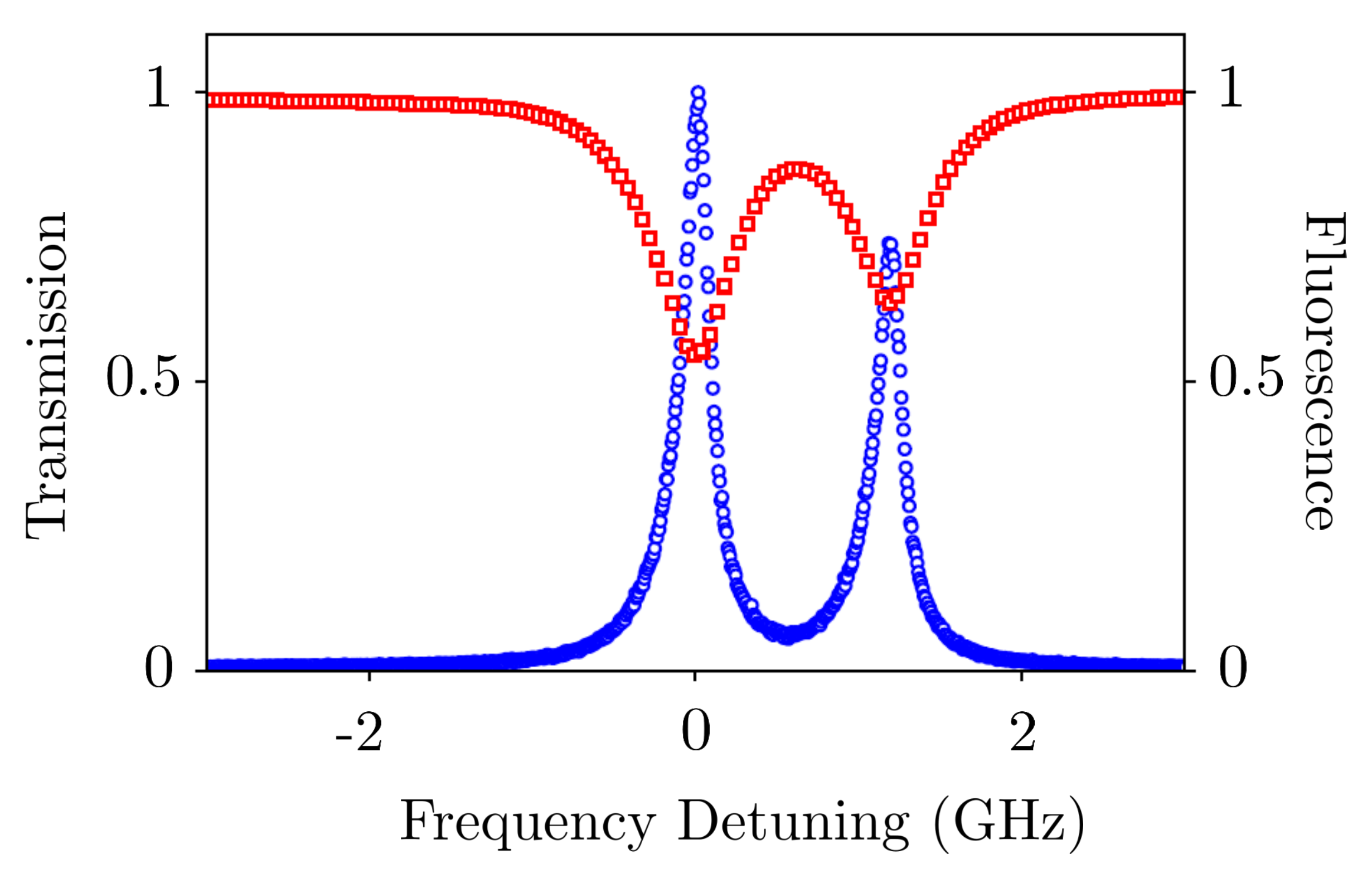}
\caption{Spectra of the D1 line of cesium acquired using the nano-cell described in the main text, heated at $174^{\circ}$C. Cell thickness: $L=447$\,nm. (a) In blue circles (red squares) : Normalized fluorescence spectrum collected at 90$^{\circ}$ from laser propagation (Absorption spectrum). Horizontal axe: laser frequency detuning with respect to the $F=4 \rightarrow F'=3$ hyperfine transition of the $6S_{1/2}$ ground state. The peak centered around $1$\,GHz corresponds to the $F=4 \rightarrow F'=4$ transition. The data are binned 10 times by steps of $2$\,MHz. (b) Fluorescence spectrum from the wedge versus the laser frequency detuning, acquired at $90^{\circ}$ from the incoming beam. The data are binned by steps of $2$\,MHz. }
\label{fig:Fig5}
\end{figure}

\section{EXAMPLE OF SPECTRA}

Finally we present examples of spectroscopic data obtained with the cell described above.
The cell was enclosed in an oven with small apertures to let a laser beam propagate through the wedge and to collect the fluorescence light at $90^{\circ}$ from the laser beam. With this set-up we acquired absorption and fluorescence spectra shown in Fig.~\ref{fig:Fig5}. These spectra were obtained by scanning the frequency of a laser diode around the D1 line of Cs at $\lambda=894$\,nm. In practice one should pay attention to maintain the temperature of the wedge $30^{\circ}$C higher that the temperature of the Cesium reservoir to avoid Cesium to condense on the wedge when the temperature is ramped down below the melting point. The spectra of Fig.~\ref{fig:Fig5} were obtained at a cell temperature of $174^{\circ}$C (temperature of the reservoir, corresponding to an atomic density $N=2.6 \times 10^{15}$\,cm$^{-3}$). The laser was focused onto the thin wedge at a point where the local thickness is $L=\lambda/2=447$\,nm. The laser power was $\sim 1\,\mu$W and the laser spot at $1/e^2$ radius was $40\,\mu$m. The transmitted light was collected onto a standard Si photodiode. We simultaneously acquired the fluorescence spectrum using a fibred avalanche photodiode in photon counting mode. The overall detection efficiency was $\sim1.5\times10^{-3}$, accounting for the solid angle collection efficiency ($7.5\times10^{-3}$), the fiber coupling ($50\%$) and the photodiode quantum efficiency ($40\%$). The possibility to acquire data on a zero background may improve signal-to-noise ratio in comparison with transmission spectra. Both spectra in fluorescence and transmission exhibit a narrow lineshape (smaller than Doppler width), characteristic of the Dicke narrowing observed in nano-cells\,\cite{Dutier}.

\section{CONCLUSION}

In conclusion, we have built a new generation of nano-cells that presents many advantages. The cell parts are made of Borofloat glass that allows a very good polishing and therefore a very small surface roughness. It also allows to connect all parts of the cell with optical bonding resulting in a monolithic cell that makes no use of any glue. It therefore avoids any degasing issues for future spectroscopic studies. For the surfaces in contact with the atoms in the nano-wedge, a super-polish is obtained and the surface roughness is less than 1{\AA} rms. The wedge has been realized and characterized using interferometric techniques and the thickness of the cell ranges from exactly zero (optical contact) to $900$\,nm. Few-nanometer thick atomic vapors have never been probed due to the way nano-cells were closed prior to this work. The type of cells presented here opens up access for the exploration of new mesoscopic systems. The spectra measured using these cells are promising for future investigation of atom-surface interactions. Our fabrication technique allows in principle to cover the cell surfaces with a dielectric layer to increase the reflectivity and hence the dipole-dipole collective effects. The fabrication is theoretically not restricted to glass and testing it with other materials could provide more data on unknown atom-surface potentials. Finally, such cells could be used as in Ref.~\cite{Sandogdhar2016} to perform molecular spectroscopy  with various confinement depths.

\section{ACKNOWLEDGMENTS}

We thank T.~Pfau for fruitful discussions and J.~Moreau for helping us analyze the AFM images. T.~Peyrot is supported by the DGA-DSTL fellowship 2015600028. We also acknowledge   financial support from CNRS, EPSRC (grant EP/R002061/1) and Durham University.



\begin{thebibliography}{80}
\bibitem{Polishing}
The fluctuations average differently for a super-polish and a standard polish. We attribute this behaviour to the fact that, following the polishing procedures, the spatial frequency distributions of surface height fluctuations are intrinsically different.
\end{thebibliography}
\end{document}